\documentclass[twocolumn,apl,amsmath,amssymb,showpacs,superscriptaddress]{revtex4-2}
\usepackage{epsf} 
\usepackage{graphicx}
\usepackage{color}
\usepackage{soul}
\usepackage{gensymb}
\usepackage{sidecap}
\usepackage{amsmath}
\usepackage{mathtools}
\usepackage{float}
 \usepackage{multirow}
\usepackage[hidelinks,colorlinks=true,linkcolor=blue,citecolor=blue]{hyperref}
\DeclareUnicodeCharacter{2212}{\textendash}

\begin{document}
\title{Improved electrochemical performance of NASICON type Na$_{3}$V$_{2-x}$Co$_x$(PO$_{4}$)$_{3}$/C ($x=$ 0--0.15) cathode for high rate and stable sodium-ion batteries}
\author{Simranjot K. Sapra}
\affiliation{Department of Physics, Indian Institute of Technology Delhi, Hauz Khas, New Delhi-110016, India}
\affiliation{International College of Semiconductor Technology, National Yang Ming Chiao Tung University, 1001 University Road, Hsinchu 30010, Taiwan}
\affiliation{Department of Materials Science and Engineering, National Yang Ming Chiao Tung University, 1001 University Road, Hsinchu 30010, Taiwan}
\author{Jeng-Kuei Chang}
\email{jkchang@nycu.edu.tw}
\affiliation{International College of Semiconductor Technology, National Yang Ming Chiao Tung University, 1001 University Road, Hsinchu 30010, Taiwan}
\affiliation{Department of Materials Science and Engineering, National Yang Ming Chiao Tung University, 1001 University Road, Hsinchu 30010, Taiwan}
\affiliation{Department of Chemical Engineering, Chung Yuan Christian University, 200 Chung Pei Road, Taoyuan, 32023 Taiwan}
\author{Rajendra S. Dhaka}
\email{rsdhaka@physics.iitd.ac.in}
\affiliation{Department of Physics, Indian Institute of Technology Delhi, Hauz Khas, New Delhi-110016, India}

\date{\today}      
	
\begin{abstract}
In recent years, the Na-ion SuperIonic CONductor (NASICON) based polyanionics are considered the pertinent cathode materials in sodium-ion batteries due to their 3D open framework, which can accommodate a wide range of Na content and can offer high ionic conductivity with great structural stability. However, owing to the inferior electronic conductivity, these materials suffer from unappealing rate capability and cyclic stability for practical applications. Therefore, in this work we investigate the effect of Co substitution at V site on the electrochemical performance and diffusion kinetics of Na$_{3}$V$_{2-x}$Co$_x$(PO$_{4}$)$_{3}$/C ($x=$ 0--0.15) cathodes. All the samples are characterized through Rietveld refinement of the x-ray diffraction patterns, Raman spectroscopy, transmission electron microscopy, etc. We demonstrate improved electrochemical performance for the $x=$ 0.05 electrode with reversible capacity of 105 mAh g$^{-1}$ at 0.1 C. Interestingly, the specific capacity of 80 mAh g$^{-1}$ is achieved at 10 C with retention of about 92\% after 500 cycles and 79.5\% after 1500 cycles and having nearly 100\% Coulombic efficiency. The extracted diffusion coefficient values through galvanostatic intermittent titration technique and cyclic voltammetry are found to be in the range of 10$^{-9}$--10$^{-11}$ cm$^{2}$ s$^{-1}$. The postmortem studies show the excellent structural and morphological stability after testing for 500 cycles at 10 C. Our study reveals the role of optimal dopant of Co$^{3+}$ ions at V site to improve the cyclic stability at high current rate. 
\end{abstract}

	\maketitle
	\section{\noindent ~Introduction}
	
Sodium-ion batteries (SIBs) have emerged as forerunner technologies in the energy storage sector, attributed to the abundant resources, wide availability, and cost-effectiveness of required raw materials \cite{Hwang_CSR_2017, Delmas_AEM_2018, Eftekhari_JPS_2018, Slater_AFM_2013, Ellis_COSSM_2012, Yabuuchi_CR_2014, Dwivedi_AEM_2021, Manish_CEJ_2023}. In recent years, a lot of research is being conducted on the development of sodium-based electrode materials and electrolytes, considering the similar intercalation chemistry and structural properties with lithium battery technology \cite{Sun_AEM_2019, Sapra_Wiley_2021, Soundharrajan_JMCA_2022}. However, the exploration of the high energy density electrodes with long cycle life and excellent rate performance remains censorious. Owing to the larger ionic radius (1.02 {\AA} for Na vs. 0.76 {\AA} for Li) and larger volume changes upon Na-insertion and extraction, structural stability poses another challenge, which affects the electrochemical performance of the SIBs \cite{Xu_AEM_2018,Peters_EES_2016}. Therefore, the large-scale production of SIBs seeks attention to bring them into the commercial layout for the practical use where the structural and electrochemical stability of cathode materials play a crucial role for high energy density and long cycle life \cite{Deng_AEM_2018, Rudola_JMCA_2021, Roberts_NSA_2018, Tang_COCE_2015, Pan_EES_2013}. A variety of cathode materials like layered oxide cathodes, honeycomb materials and polyanionic compounds have been explored \cite{Pati_JMCA_2022, Saroha_ACSOmega_2019, Chandra_EA_2020, Pati_JPS_24}. In this line, Sodium Super Ionic CONductors (NASICON) structures with the highly covalent three-dimensional open framework for the facile diffusion of the sodium ions upon intercalation/de-intercalation are widely explored as potential cathode materials for SIBs \cite{Goodenough_MRB_1976, Jian_AM_2017}. Amongst the various NASICON compounds, Na$_{3}$V$_{2}$(PO$_{4}$)$_{3}$ (NVP) exhibits excellent structural stability, attractive operating potential (V$^{4+}$/V$^{3+}$ at 3.4~V vs Na$^{+}$/Na), moderate specific theoretical capacity of 117.6 mAhg$^{-1}$ with high thermodynamic stability and rapid diffusion channels, providing excellent ionic conductivity, which makes the NVP pertinent for practical utilization \cite{Ouyang_NCOM_2021, Zheng_JEC_2018, Wang_AEM_2020, Rajagopalan_ESM_2021, Song_JMCA_2014}. However, the low intrinsic electronic conductivity, arising from the distorted VO$_{6}$ octahedra due to the presence of large anionic group skeletons hinders the superior electrochemical performance \cite{Chen_AM_2017}. To alleviate this issue, various modifications are incorporated into the structure: (i) introduction of the alien ions into the bulk structure and (ii) coating some conductive carbon matrices onto the material \cite{Zhang_JALCOM_2021,Lv_CNM_2019}. The carbon encapsulation can improve the chemical stability and high-rate performance; however, the energy density is reduced due to the large amount of inactive carbon and smaller tap densities. On the other side, the systematic introduction of vacancies can alter the electronic and crystal structures, modify the bottleneck of the intrinsic lattice and reduce the diffusion barriers, thereby boosting the Na-ion diffusion across the bulk of the electrode material \cite{Li_AEM_2020, Xiao_SMALL_2023}.  

Recent studies showed that the effective substitution of the Ni$^{2+}$, Cr$^{3+}$, Mg$^{2+}$, Mn$^{2+}$, Al$^{3+}$, Ce$^{4+}$, Zr$^{4+}$ and Mo$^{6+}$ at the V site yields enhanced electrochemical performance \cite{Li_JMCA_2015, Li_AMI_2016, Li_JMCA_2018, Ghosh_AES_2022, Wu_AEM_2021, Zakharkin_JPS_2020, Aragon_CEC_2015, Lavela_JPS_2021, Zheng_ECA_2017, Chen_JPS_2018}. Interestingly, the Co$^{3+}$ doping in layered oxide  Na$_{0.44}$Mn$_{0.09}$Co$_{0.01}$O$_{2}$ having P2-type tunnel structures increases the Mn$^{4+}$/Mn$^{3+}$ ratio along with the suppression of Jahn-Teller distortion resulting in the structural stability \cite{Oz_ACS_Omega_2023}. Also, Co$^{3+}$ doping prevents the cationic disorder in LiNi$_{0.94}$Co$_{0.06}$O$_{2}$ with good thermal and structural stability and improved electrochemical performance for Li-ion, Na-ion and supercapacitor applications \cite{Karuppiah_IJES_2020}. Lithium and sodium-based hybrid layered oxide, Na$_{2/3}$Li$_{1/3}$Mn$_{0.95}$Co$_{0.05}$O$_{2}$ with 5\% Co concentration at Mn site showed reversible capacity of 112 mAh/g with the capacity retention of 97.7\% after 100 charge-discharge cycles at 0.1~C \cite{Hoa_JEC_2023}. There are also many reports on the development of low cobalt active materials due to their high specific capacity and energy density, as summarized in \cite{Mallick_JMCA_2023}. It has been reported that the Co substitution exhibits excellent chemical stability, structural stability and dissolution stability in layered oxides owing to the large octahedral site and stabilization energy as compared to Ni/Mn doping \cite{Manthiram_Nature_Com_2020}. However, the effect of Co substitution on the electrochemical performance has not been explored in NASICON type polyanionic Na$_{3}$V$_{2}$(PO$_{4}$)$_{3}$ cathode. 

Therefore, inspired from the positive attributes of the cobalt dopant, in this paper we choose to investigate the electrochemical diffusion kinetics of Na$_{3}$V$_{2-x}$Co$_{x}$(PO$_{4}$)$_{3}$/C ($x=$ 0--0.15) cathodes for the sodium-ion batteries. Considering the environmental precariousness of the cobalt, low concentrations are chosen to tune the electrodes for high specific capacity and energy density. Also, particle size reduction is achieved through the high energy ball milling to shorten the ion diffusion length and increase interfacial contacts. The comprehensive effects of cobalt substitution and sodium diffusion kinetics are studied using cyclic voltammetry (CV), galvanostatic cycling, electrochemical impedance spectroscopy (EIS), and galvanostatic intermittent titration technique (GITT). The inclusion of Co dopant into the bulk of the NVP electrodes has improved sodium ion diffusion and mobility, resulting in an improved electrochemical performance at higher rates, serving suitable candidates for the fast-charging and highly stable sodium-ion batteries. 
	
\section{\noindent ~Experimental Section}
\noindent 2.1 \textit{~Synthesis of materials}:
Na$_3$V$_{2-x}$Co$_{x}$(PO$_{4}$)$_{3}$/C ($x=$ 0, 0.05, 0.10, 0.15) samples were prepared using the facile ball-milling assisted solid-state reaction route. The chemicals, i.e., sodium carbonate (Na$_{2}$CO$_{3}$, Alfa-Aesar, $\mathrm{>}$99\%), ammonium dihyrdrogen phosphate (NH$_{4}$H$_{2}$PO$_{4}$, Alfa-Aesar, $\mathrm{>}$99\%), cobalt oxide (Co$_{3}$O$_{4}$, Alfa-Aesar, $\mathrm{>}$99.9\%) and divanadium pentaoxide(V$_{2}$O$_{5}$, Alfa-Aesar, $\mathrm{>}$99\%) were used according to the stoichiometric ratio. Excess 5 wt.\% Na$_{2}$CO$_{3}$ was added to compensate for Na loss during heat treatment. 5 wt.\% of acetylene black was used as a conducting and reducing agent for the reduction of V$^{+5}$ to V$^{+3}$. The stoichiometric amount of the above precursors were dissolved in ethanol (99.9\%) and were ball milled for 24 hrs. The resultant mixture was dried at 80\degree C for 12 hrs in an oven. After grinding the powder for about 1 hr, the powder was calcined in a tube furnace at 350\degree C for 4 hrs and 800\degree C for 8 hrs in an argon hydrogen atmosphere (9:1) to obtain the final samples. The samples are abbreviated as NVP, NVP-Co5, NVP-Co10 and NVP-Co15, respectively. \\
\noindent 2.2 \textit{ Fabrication of electrodes and cell assembly}:
A cathode slurry was prepared by mixing 80 wt.\% active powder, 10 wt.\% Super P, and 10 wt.\% PVDF (Polyvinylidene fluoride) in N-methyl-2-pyrrolidone solution. This slurry was pasted onto the battery grade Al foil and vacuum-dried at 90\degree C for 10 hrs. The electrode was punched based on the required dimensions of CR2032 and the active mass loading $\mathrm{\sim}$2.5 mg cm$^{-2}$ was obtained. The cells were assembled using a thin Na metal foil as the negative electrode, Glass Fiber Separator (Advantec) as the separator and 1 M NaClO$_{4}$ in ethylene carbonate (EC) and polypropylene carbonate (PC) with 5 wt.\% fluoroethylene carbonate (FEC) additive as the electrolyte in an argon-filled glove box (Innovation Technology Co. Ltd.), where both the moisture and oxygen content levels were maintained at $\mathrm{\sim}$0.3 ppm.

	\begin{figure*}
	\includegraphics[width=17.5cm,height=14.5cm]{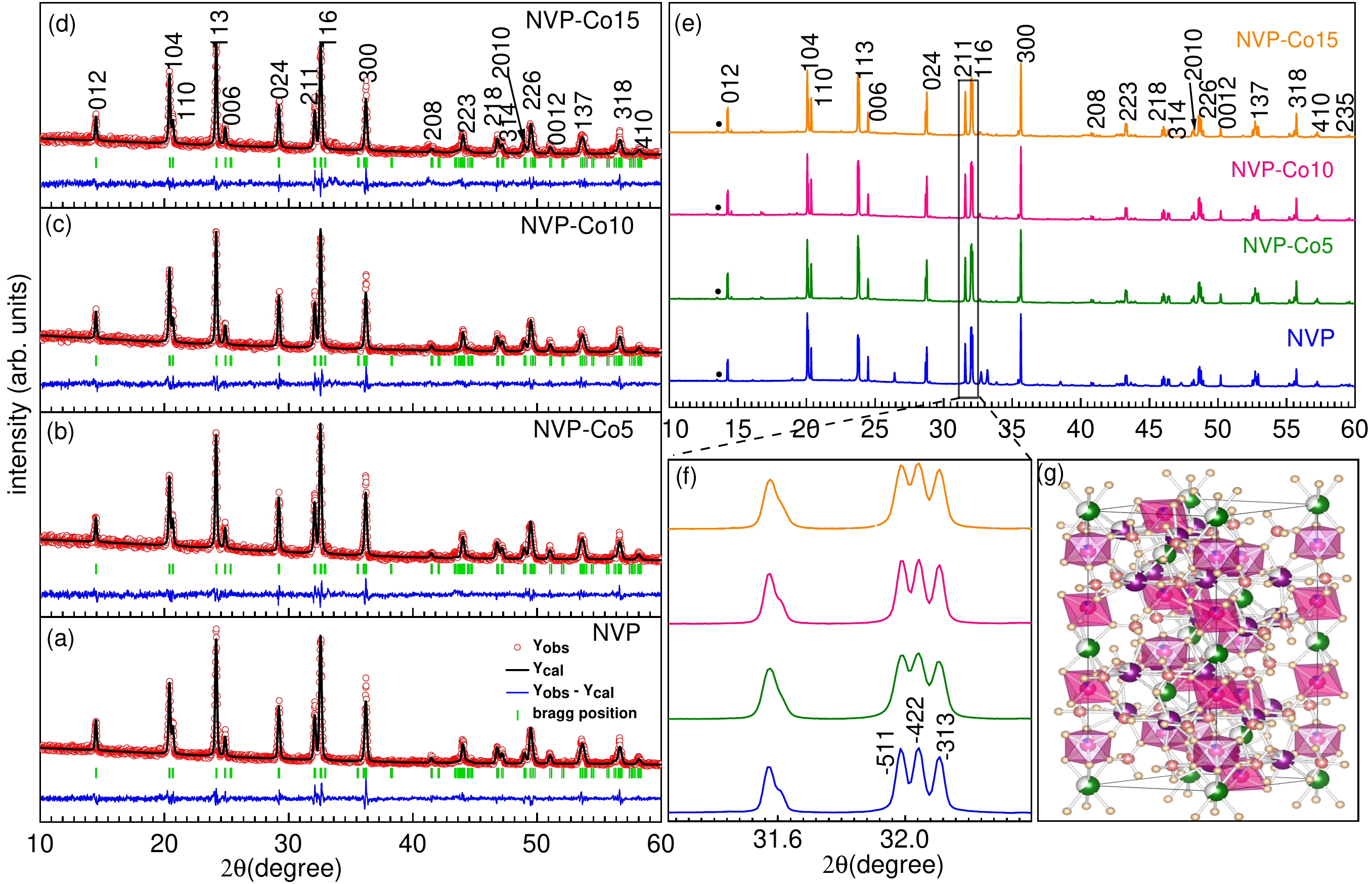}
		\caption{The powder x-ray diffraction (XRD) patterns (red), the Rietveld refined profiles (black), the fitting residue (blue) and the Bragg positions (green bars) (a-d) and (e) the Synchrotron XRD patterns for the Na$_{3}$V$_{2-x}$Co$_{x}$(PO$_{4}$)$_{3}$ ($x=$ 0, 0.05, 0.1, 0.15) samples. (f) The magnified Synchrotron XRD pattern of Na$_{3}$V$_{2-x}$Co$_{x}$(PO$_{4}$)$_{3}$ in the 2$\theta$ range of 31.25 -- 32.39 $\degree$ and (g) the crystal structure of NVP-Co5 sample where V, Na1, Na2, P, O, C, Co are depicted by the magenta, pink, purple, orange, yellow and blue, drawn using the VESTA software.}
		\label{XRD-1}
	\end{figure*}
	
	\noindent 2.3. \textit{Material characterization}: 
	Synchrotron X-ray diffraction (SXRD) was conducted in the range of 2$\theta$ = 10$\mathrm{{}^\circ}$--80$\mathrm{{}^\circ}$ using a 0.3~mm diameter capillary (wavelength of 0.61992~{\AA}), at National Synchrotron Radiation Research Centre (NSRRC, Hsinchu). We also used the laboratory based X-ray diffraction with Cu K$_{\alpha}$ radiation with a wavelength of 1.5418~{\AA} to investigate the crystal structure. The sample morphology was examined with the field emission scanning electron microscopy JEOL-FESEM (EVO18), and transmission electron microscopy (TEM, JEOL-JEM-1400). Energy-dispersive X-ray spectroscopy (EDS) was employed to probe the elemental mapping. The Raman modes were studied with the Micro Raman Spectrometer (Renishaw) with a probing wavelength of 514 nm. The vibrational modes were measured with the Fourier Transmission Infrared Spectroscopy (FTIR, Thermo Nicolet-IS-50) in the spectral range of 500-2000 cm${}^{\mathrm{-}}$${}^{1\ }$by preparing a 13~mm diameter pellet of the mixture of the sample with the potassium bromide KBr. X-ray photoelectron spectroscopy (XPS from Thermo VG-Scientific/Sigma Probe) measurements were performed with Al K\textit{${}_{\alpha \ }$}radiation (1486.6 eV) X-ray source, and the analyzer pass energy of 20~eV and step size of 0.1~eV. The data fitting was conducted using the XPSPEAK 4.1 software using C 1$s$ binding energy (284.8~eV) as a reference to examine the oxidation state. The elemental compositions of the samples were determined through Inductively Coupled Plasma-Optical Emission Spectrometer (ICP-OES, American Agilent 725). 
	
\noindent 2.4. \textit{Electrochemical testing}: The charge and discharge performances of the fabricated cells were evaluated between 2.0 and 4.3 V with various currents using Neware BTS8.0 battery tester at room temperature. The cyclic voltammetry (CV) was studied using the BT-LAB V1.64 potentiostat at a sweep rate of 0.1 - 1.0 mV/s. The electrochemical impedance spectroscopy (EIS) measurements were conducted in the frequency range of 100 kHz to 10 mHz with an AC amplitude of 10 mV using VSP300 potentiostat. The galvanostatic intermittent titration technique (GITT) measurements were conducted using the Arbin battery cycler with a pulse time of 20 minutes at a 0.1~C rate, followed by a relaxation of 3 hrs. 
	
\section{\noindent ~Results and discussion}

\begin{figure}
\includegraphics[width=\linewidth]{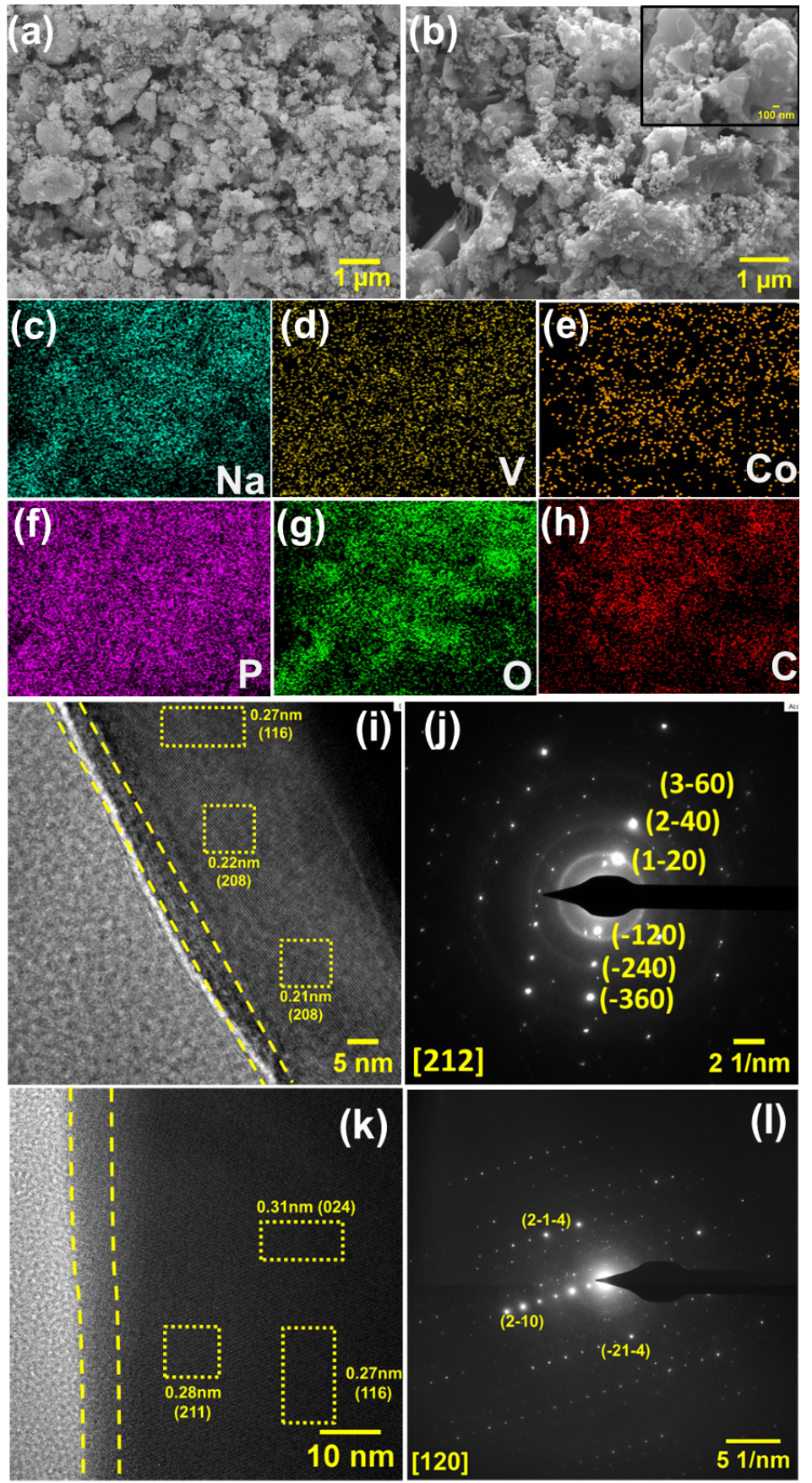}
\caption{The FE-SEM image of the NVP and NVP-Co5 sample (a, b); the EDS mapping of all the elements in NVP-Co5 in (c--h); the HR-TEM diffraction fringes and SAED pattern for NVP (i-j) and NVP-Co5 (k-l), respectively.}
\label{fesem-tem}
\end{figure}
	
First we present the X-ray diffraction (XRD) patterns of Co substituted Na$_{3}$V$_{2-x}$Co$_{x}$(PO$_{4}$)$_{3}$/C ($x=$ 0--0.15) samples in Figs.~\ref{XRD-1}(a--d), which confirm the rhombohedral ($R\bar{3}c$) symmetry in the NASICON structure. The observed diffraction peaks without any impurity demonstrate the successful substitution of Co in the host crystal lattice of pristine NVP sample. Further, we perform the Rietveld refinement of XRD patterns using the Pseudo-Voigt line function and six-coefficient polynomial background, as displayed in Figs.~\ref{XRD-1} (a-d). Through the Rietveld refinement, we extract the lattice parameters and cell volume, which are presented in Table 1 of Supplementary Information along with crystallite size, different bond lengths and R-factors. The lattice parameters obtained show small variations with the substitution, owing to the similar ionic radii of Co$^{3+}$ (0.61 \AA) and V$^{3+}$ (0.64 \AA). A slight decrease observed in $c-$parameter for NVP-Co10 and NVP-Co15 samples may be attributed to the increase in the population of Na1 sites, resulting in a reduction in electrostatic repulsion between the oxygen layers that are perpendicular to the $c-$axis of the unit cell \cite{Ghosh_AEM_2020, Ghosh_AFM_2021, Ghosh_AES_2022}. Similarly, lattice constant $a$ with the tiny variation upon the substitution (8.59 \AA) confirms the identical [O3NaO3(V/Co)O3NaO3] column size \cite{Ghosh_AEM_2020}. The detailed bond lengths for the atoms from the Bond Valence Sum (BVS) calculations and the refined atomic positions and occupancies for all the samples are also presented in Tables 1--4 of Supplementary Information. 
	
\begin{figure*}
\includegraphics[width=\linewidth]{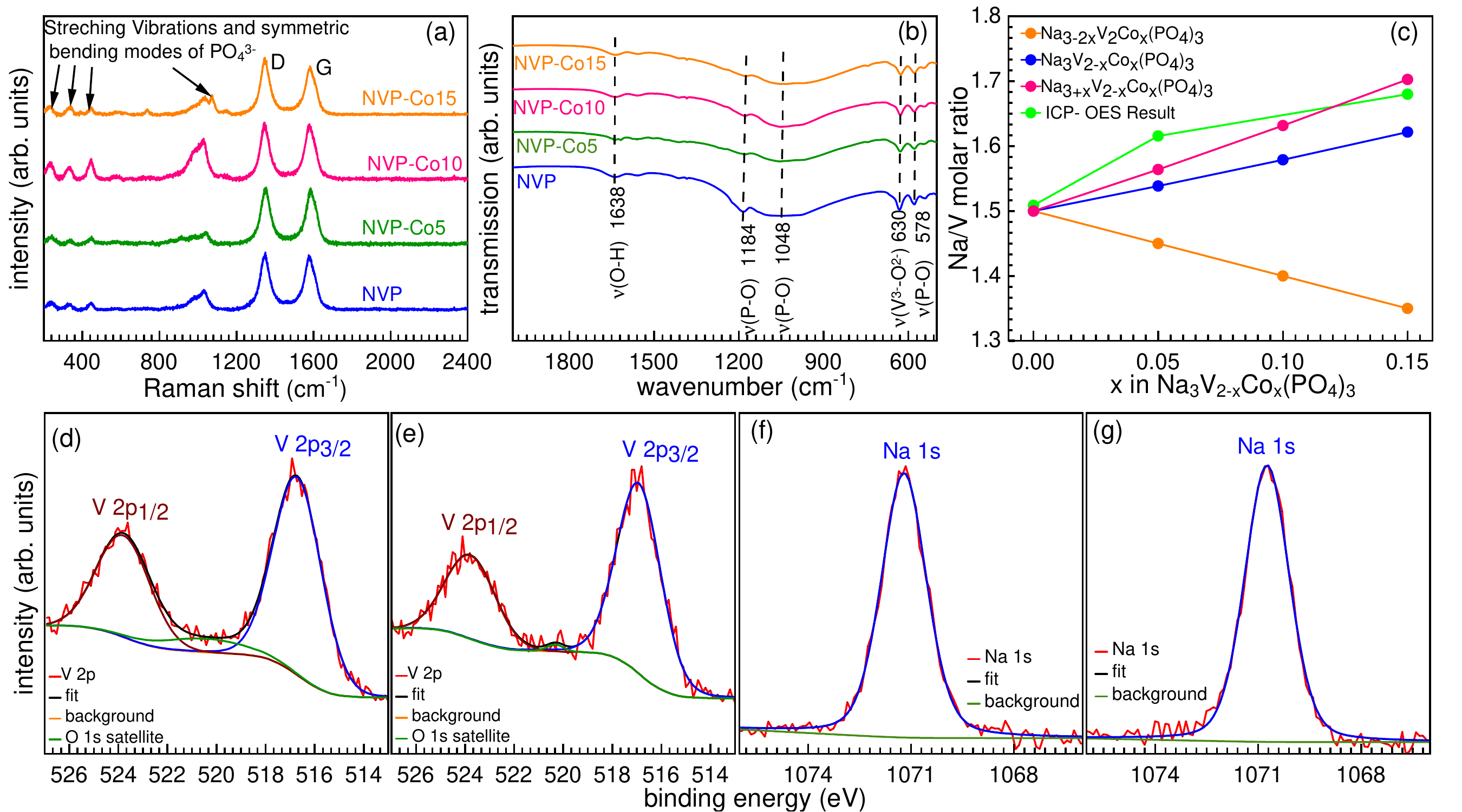}
\caption{The Raman spectra (a) and the FTIR spectra (b); the Na/V ratio from the ICP-OES results (c) for the Na$_{3}$V$_{2-x}$Co$_{x}$(PO$_{4}$)$_{3}$/C ($x=$ 0, 0.05, 0.1, 0.15) powder samples; The V 2$p$ core level spectra for the NVP (d) and NVP-Co5 (e) samples. The Na 1$s$ core level spectra of the NVP (f) and NVP-Co5 (g) samples, respectively.}
\label{FTIR-RAMAN}
\end{figure*}
	
Moreover, we use high-resolution synchrotron based X-ray diffraction (SXRD), which confirms the high crystallinity and the patterns are mostly indexed using the rhombohedral (space group $R\bar{3}c$) unit cell \cite{Wang_AEM_2020, Rajagopalan_ESM_2021, Song_JMCA_2014}, as shown in Fig.~\ref{XRD-1}(e). However, Gaubicher {\it et al.} reported the existence of monoclinic distortion at room temperature due to Na$^{+}$ ion ordering between Na1 and Na2 crystallographic sites or vacancy defects (due to composition variations during synthesis) within the V$_{2}$(PO$_{4}$)$_{3}$ framework, leading to reduction in the crystal symmetry ($C2/c$ space group) \cite{Gaubicher_CM_2000,Zou_AFM_2021}. This means the NVP unit cell goes from ordered ($C2/c$) at room temperature to the disordered ($R\bar{3}c$) symmetry at high temperature of around 200$\degree$C; however, the general skeleton of VO$_{6}$/CoO$_{6}$ octahedra and PO$_{4}$ tetrahedra forming a lantern unit is preserved in both the symmetries \cite{Chotard_CM_2015, Lalere_JPS_2014}. In this context, room temperature high-resolution SXRD measurements for all the samples aid in distinguishing monoclinic distortions in these complex crystal structures, which are not clearly visible by the laboratory X-ray diffraction patterns. We observe the splitting of (116) and (211) peak reflections of the unit cell, as shown in magnified view in Fig.~\ref{XRD-1}(f). The (116) reflection splits into (-511), (-422) and (-313) reflections, which are allowed in $C2/c$ unit cells, confirm the monoclinic distortions in the crystal framework \cite{Park_ChemMater_2021}. We also observe a few weak superstructure reflections at the lower angles, as marked by a black circle in Fig.~\ref{XRD-1}(e), which also emerge from the occupancies and positions of Na$^{+}$ ions in the NASICON framework, referred to as incommensurate modulated structure, for which there are no detailed studies reported at room temperature \cite{Park_ChemMater_2021}. Hence,  the XRD patterns, shown in Figs.~\ref{XRD-1}(a--d) are refined using the rhombohedral $R\bar{3}c$ symmetry for the sake of comparison of all the samples and due to the large number of Wyckoff sites available in the lower symmetry, the use of $C2/c$ does not show significant improvement in the refinement procedures for the presented samples \cite{Ghosh_AFM_2021}. We find that the crystal structure of Co substituted NVP samples consist of a three-dimensional framework of two octahedra VO$_{6}$/CoO$_{6}$ combined with three tetrahedra PO$_{4}$ forming a lantern unit, stacked along the (001) plane, as shown in Fig.~\ref{XRD-1}(g) for the NVP-Co5 sample. Each lantern unit combines with six other lantern units, forming a large interstitial space that can contain 0 to 5 alkali ions, depending on the oxidation state of the metal and phosphorous, which can generate different ion conduction pathways \cite{Jian_AM_2017}. The Na ions are located at the two sites, 6b (Na1) and 18e (Na2), with an occupancy of 1 and 0.67, respectively, in this crystal framework. Interestingly, both the Na1 and Na2 sites  garner significant attention for the battery performance as 2 Na$^{+}$ ions easily intercalate from Na2 sites, while the Na$^{+}$ ion at Na1 site is incapable of participating in the electrochemical reactions \cite{Song_JMCA_2014}. 
 
The field emission scanning electron microscopy (FE-SEM) measurements are employed to confirm the morphology and microstructure of the samples. The morphology indicates the presence of irregular secondary particles in the 1$\mu$m range with a similar configuration, composed of primary nanoscale particles, as shown in Figs.~\ref{fesem-tem}(a, b) for NVP and NVP-Co5 samples, respectively, arising from the inevitable agglomeration during the high-temperature calcination. It also demonstrates that the substitution of the Co$^{3+}$ ion does not affect the particle growth and morphology, thereby maintaining the structural integrity of the NVP crystal framework. The particle size is also confirmed from the dynamic light scattering (DLS) measurements, which show that particle size is in $\mu$m regime for all the samples, shown in Supplementary Information. The EDS mapping of all the elements Na, O, V, Co, P, and C of NVP-Co5 sample confirm the successful incorporation of the Co$^{3+}$ ion and uniform atomic distribution across the selected area, as shown in Figs.~\ref{fesem-tem}(c--h). The detailed microstructures for the NVP and NVP-Co5 samples are also investigated with the high-resolution TEM, which shows the clear lattice fringes. The NVP sample exhibits $d-$spacing of 0.21 nm and 0.27 nm, corresponding to the (208) and (116) planes, respectively, as shown in Fig.~\ref{fesem-tem}(i). The $d-$spacing of NVP-Co5 sample is found to be 0.28 nm and 0.27 nm, corresponding to the (211) and (116) planes, as shown in Fig.~\ref{fesem-tem}(k). The particle is also surrounded by the thin amorphous carbon layer, which further aids in increasing the extrinsic electrical conductivity of the active material by enabling the movement of the Na ions and electrons. The selected area electron diffraction (SAED) patterns, in Figs.~\ref{fesem-tem}(j, l) exhibit bright spots with strong contrast corresponding to the different crystallographic planes, which demonstrate the high crystallinity of both the samples, and the electron diffraction data are found to be consistent with the aforementioned XRD study. 
	
The Raman spectra are collected to reveal the presence of carbon layers and vibrational modes, as shown in Fig.~\ref{FTIR-RAMAN}(a). The bands in the range of 300-1200 cm$^{-1}$ arise from the stretching vibrations and symmetric bending modes of the PO$_{4}$$^{3-}$ anionic groups, confirming the indistinguishable crystal structure upon the substitution \cite{Manish_PRB_24}. The presence of carbon is signaled by the two characteristic bands, D (1346 cm$^{-1}$) and G (1600 cm$^{-1}$). The D band corresponds to the disordered carbon, which is attributed to the lack of long-range translational symmetry in amorphous carbons and the G band corresponds to the graphitic carbon, also called Raman-allowed E$_{2g}$ mode resulting from the in-plane displacement of carbon atoms in the graphene sheets \cite{Ferrari_PRB_2000}. The intensity ratio of D and G bands specifies the degree of graphitization of the material, which is obtained as 1.04, 0.99, 1.00, and 1.18 for the NVP, NVP-Co5, NVP-Co10, and NVP-Co15 samples, respectively \cite{Li_JMCA_2015}. The values of the I\textit{$_{D}$}/I\textit{$_{G}$} ratio demonstrate that coatings comprise predominantly sp$^{2}$ carbon, has considerable electrical conductivity, boosting the dispersion of the electrons over the bulk of the electrode. It also confirms that the substitution of Co does not affect the intensity ratio significantly. The chemical bonds associated with the NASICON structure are further investigated through the FTIR measurements, as shown in Fig.~\ref{FTIR-RAMAN}(b). The stretching vibrations of the V$^{3+}$- O$^{2-}$ bonds in the VO$_{6}$ octahedra is identified at 630 cm$^{-1}$. The vibrations of the P-O bonds in the PO$_{4 }$ tetrahedra are observed at 578 cm$^{-1}$, 1048 cm$^{-1}$ and 1184 cm$^{-1}$ \cite{Zheng_ECA_2017}. The bond arising from the hydroxide groups is visible at 1638 cm$^{-1}$. The spectrum illuminates the uniform crystal structure for all the samples without significant variations in the bond lengths.  
	
The explicit elemental dopant site is further affirmed by the ICP-OES measurements, where the Na/V molar ratio is the excellent indicator to determine the effective dopant site. According to ref.~\cite{Li_FML_2013, Li_CoM_2018}, three doping phenomena are proposed when the dopant is substituted in the NASICON structure: (a) substitution at V site with additional sodium at Na site, resulting in Na$_{3+x}$V$_{2-x}$Co$_{x}$(PO$_{4}$)$_{3}$/C composition; (b) the doping at the Na site yielding composition Na$_{3-2x}$Co$_{x}$V$_{2}$(PO$_{4}$)$_{3}$/C with the extraction of sodium from Na sites; (c) doping at the V site leading to the final composition Na$_{3}$V$_{2-x}$Co$_{x}$(PO$_{4}$)$_{3}$/C. The elemental ICP-OES results for the Na/V ratio are depicted in Fig.~\ref{FTIR-RAMAN}(c), and the trend suggests that Na concentration is increasing at the Na site along with the reduction in V concentration and our results match with the doping mechanism (a). In this fashion, there can be the possibility of co-existence of the mixed Co$^{2+}$/Co$^{3+}$ ions in the system as the initial precursor used to synthesize the material is in a mixed valence state, although the most stable state of cobalt is +3. Hence, to maintain charge neutrality, more Na ions are introduced in the lattice when mixed valence Co ion replaces the V because of the charge difference. The elemental compositions of all the samples are obtained through the molar ratios from ICP-OES results and are described in the table~I, which found to be in agreement with the expected theoretical compositions. 
 \begin{table}[h]
 	\caption{ICP-OES results for the Na$_{3}$V$_{2-x}$Co$_{x}$(PO$_{4}$)$_{3}$/C samples, described in molar ratio }
 	\begin{tabular}{p{2cm} p{1.2cm} p{1.2cm} p{1.2cm} p{1.5cm}}
 		\hline
 		composition&Na &V &Co &Na/V \\
 		\hline
 		$x=$ 0 &0.55 &0.39&0&1.51 \\
 		$x=$ 0.05 &0.56 &0.37 &0.01&1.62\\
 		$x=$ 0.15 &0.55 &0.35&0.03&1.68\\
 		\hline
 	   \label{ICP-OES}
 	\end{tabular}
 \end{table}
Further, to gain more insight into the chemical valence state of the vanadium and sodium in the pristine and substituted samples, X-ray photoelectron spectroscopy (XPS) measurements were performed and the V 2$p$ and Na 1$s$ core-levels of the NVP and NVP-Co5 samples are depicted in Figs.~\ref{FTIR-RAMAN}(d--g). We use the Voigt function for line shape to include contributions from Lorentz and Gaussian profiles and the Shirley function for the background and de-convoluted all the core-level peaks. The V 2$p_{3/2}$ and V 2$p_{1/2}$ in the NVP are located at 516.7 eV and 523.8 eV, which are close to the reported binding energy values of V$^{3+}$ \cite{Wu_MD_2020, Bi_SCE_2019}. On the effective substitution of cobalt at the V site, there is no significant change in the binding energy values of the V 2$p_{3/2}$ and V 2$p_{1/2}$ for the NVP-Co5 sample, which illustrates that the chemical environment of the V$^{3+}$ is almost unaltered upon cobalt doping \cite{Wu_MD_2020, Bi_SCE_2019}. We also observe satellite peaks of O 1$s$ at around 520 eV due to the hybridisation of V and O orbitals \cite{Wang_IC_2020, Zhang_JMCA_2018}. Furthermore, the Na 1$s$ binding energy values are found to be at 1070.9 eV and 1070.7 eV for the NVP and NVP-Co5 samples, respectively, which are very close to the reported values in \cite{Aragon_CEJ_2017} and elucidate that chemical bonding remains unchanged. Owing to the lower concentration of cobalt in the system and experimental limitation to probe the low concentration, the Co 3$d$ core-level was too noisy (not shown). The XPS results reported here are in line with our design and agree with the XRD and ICP-MS results. 
	
\begin{figure}[h]
\includegraphics[width=\linewidth]{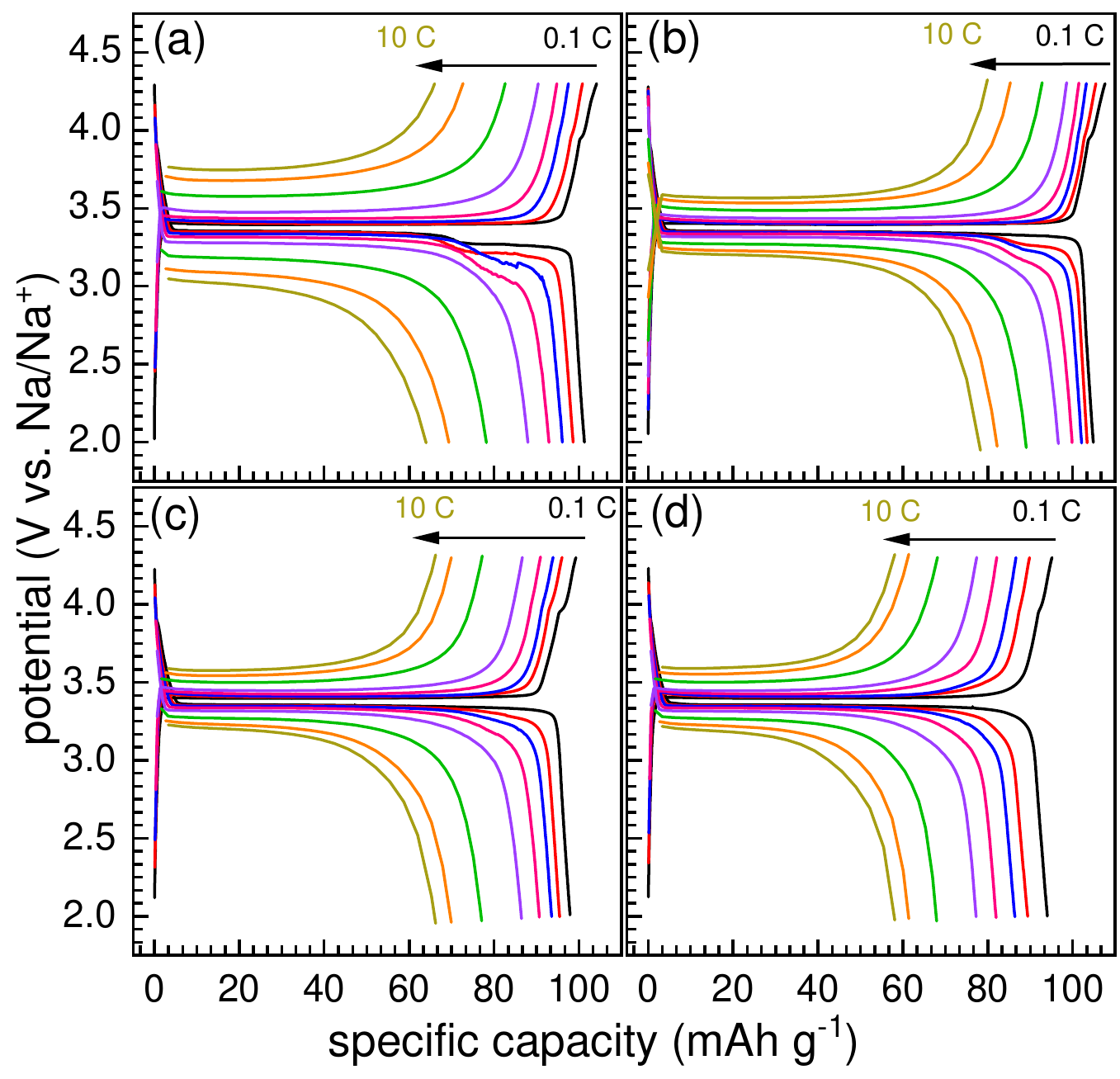}
\caption{(a--d) The galvanostatic charge-discharge characteristics at the different C-rates for all the Na$_{3}$V$_{2-x}$Co$_{x}$(PO$_{4}$)$_{3}$/C ($x=$ 0, 0.05, 0.1, 0.15) electrodes.}
\label{GCD}
\end{figure}
	
\begin{figure*}
\includegraphics[width=\linewidth]{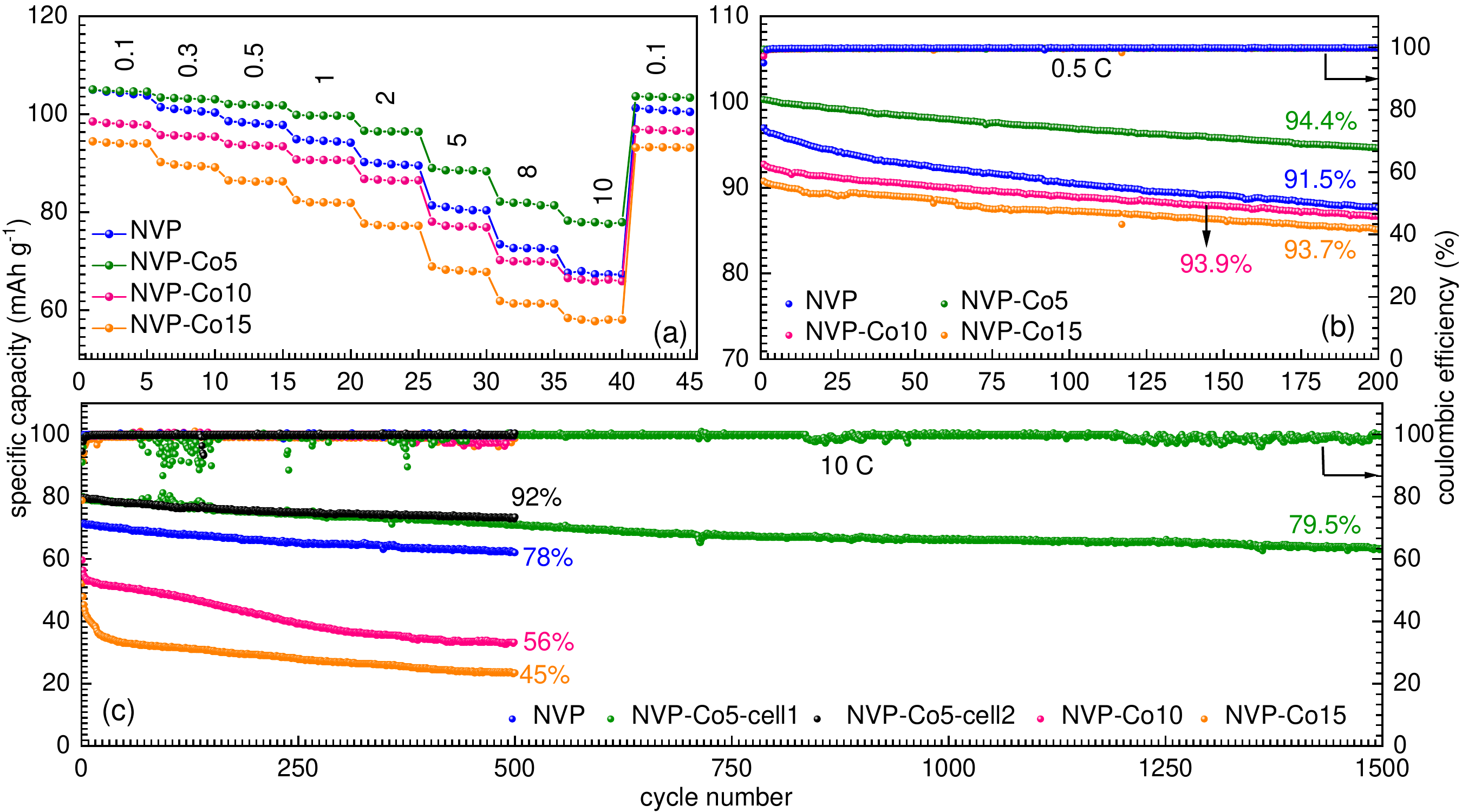}
\caption{The rate capability at the different current rates for Na$_{3}$V$_{2-x}$Co$_{x}$(PO$_{4}$)$_{3}$/C ($x=$ 0, 0.05, 0.1, 0.15) composites. The specific capacity and the Coulombic efficiency of all the cathode materials for the long cycle performance at the (b) 0.5~C for 200 cycles and (c) 10~C rates for 500 cycles, respectively. In (c) the long cycling data are also presented for another cell of best performing electrode (NVP-Co5) upto 1500 cycles.}
\label{RATE}
\end{figure*}
	
Now we focus on the electrochemical performance of all the electrodes in the half-cell configuration at room temperature and present the galvanostatic charge-discharge profiles  from 0.1 C to 10 C current rates in Fig.~\ref{GCD}. The theoretical capacities for NVP, NVP-Co5, NVP-Co10 and NVP-Co15 are 117.6, 117.46, 117.36 and 117.26 mAh g$^{-1}$, corresponding to the intercalation of two sodium ions. All the samples exhibit similar electrochemical signatures where the sodium ion extraction occurs at 3.4 V and insertion at 3.36 V, corresponding to the redox couple (V$^{3+}$/V$^{4+}$). The reaction mechanism is based on the reversible two phase change reaction, from Na$_{3}$V$_{2-x}$$^{3+}$Co$_{x}$(PO$_{4}$)$_{3}$ to Na$_{1+x}$V$_{2-2x}$$^{4+}$Co$_{x}$(PO$_{4}$)$_{3}$ with the transfer of (2$-x$) Na$^{+}$ ions and (2$-x$) electrons. Interestingly, the presence of one remaining Na$^{+}$ ion at Na(2) site in the crystal structure further lends the opportunity to access the V$^{4+}$/V$^{5+}$ redox couple at a voltage of 3.95 V, giving rise to the additional electrochemical activity at the higher voltage, as first reported in ref.~\cite{Gopalkrishnan_ChemMater_1992}. Hence, we observe a small plateau at around 3.95 V, confirming the partial oxidation of V$^{4+}$ to V$^{5+}$, as full extraction of sodium ions may distort the crystal framework. Herein, we could not observe the redox activity from Co contributing to the total electrochemical activity in this voltage window due to its higher operating voltage of Co redox couple. In addition, we also observe a splitting of the voltage plateau at 3.2 V in the discharge profile, which is associated with the rearrangement of the local crystal environment with the exchange of sodium between Na1 and Na2 sites \cite{Jiang_JMCA_2016}. The reversible discharge capacity at 0.1 C current rate for the NVP cathode is obtained around 104 mAh g$^{-1}$, which slightly increase (105 mAh g$^{-1}$) for NVP-Co5 and then start decreasing for higher Co concentration electrodes like NVP-Co10 (98 mAh g$^{-1}$) and NVP-Co15 (94 mAh g$^{-1}$). 

The impact of Co substitution on the electrochemical profiles of NVP is revealed through the rate properties and the cycling stability of all the samples, which are crucial parameters for practical applications. Fig.~\ref{RATE}(a) depicts the rate capability for all the electrodes as the current rate increases from 0.1 C to 10 C. It is observed that the NVP-Co5 cathode exhibits superior rate performance with the reversible capacity of 103.4, 102.0, 99.8, 96.6, 89.0, 82.2, and 78.2 mAh g$^{-1}$ at 0.3 C, 0.5 C, 1 C, 2 C, 5 C, 8 C and 10 C rate, respectively, in comparison to the pristine NVP, especially at higher C-rates. The obtained results convince us that cobalt doping has a significant influence on the electrochemical properties of the NVP, owing to the enhanced structural stability. However, when the Co content crosses the optimum concentration ($x=$ 0.05), the reaction kinetics is slowed down where the specific capacity and cyclic stability start declining for the NVP-Co10 and NVP-Co15 electrodes. 	
\begin{figure*}
\includegraphics[width=\linewidth]{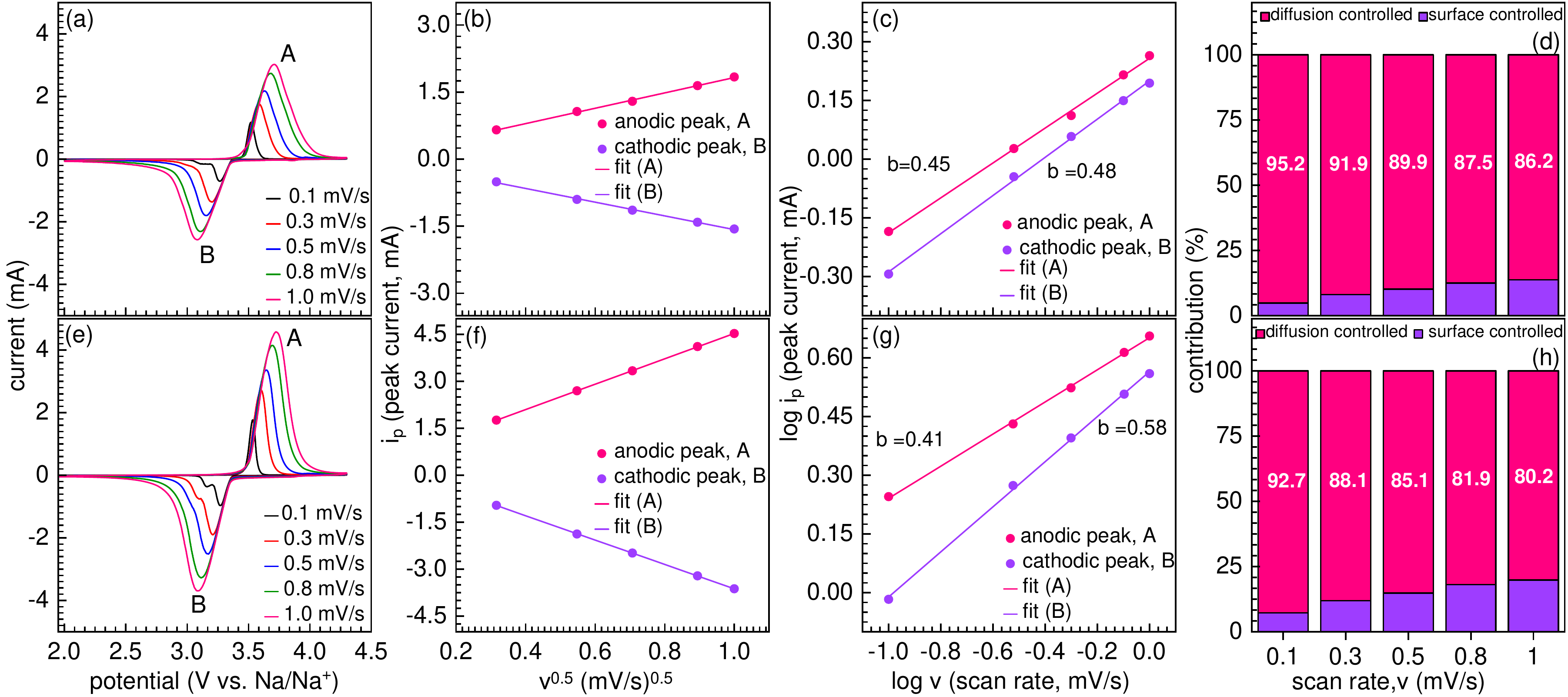}
\caption{The cyclic voltammogram at scan rates from 0.1 - 1.0 mV/s, the peak current versus scan rate, the log of the peak current versus log of scan rates, and the contribution of the capacitive and diffusive components from the total current at all scan rates, for the NVP (a--d) and NVP-Co5 (e--h) cathodes, respectively.}
\label{CV_CONTRIBUTION}
\end{figure*}
As the cycling stability is directly linked with the structural stability, the coin cells of all the samples are tested with the cycling at a low rate of 0.5 C for 200 cycles, as shown in Fig.~\ref{RATE}(b). The capacity retention of NVP, NVP-Co5, NVP-Co10 and NVP-Co15 is found to be 91.5\%, 94.4\%, 93.9\%, 93.7\%, respectively with the Coulombic efficiency of around 100\% indicating the NVP-Co5 electrode having superior cyclability. However, when the electrodes are cycled at higher rates, differences in the electrochemical performances become evident as it is hard for Na$^{+}$ ion to go into the bulk of the active particles due to the diffusion limitation and increased polarization. We tested all the samples for the long-term cycling capability for 500 cycles at a high rate of 10 C and the results are depicted in Fig.~\ref{RATE}(c). Interestingly, the outstanding retention of 92\% is observed for the NVP-Co5 cathode after 500 cycles, which is found to be much better as compare to others like the NVP, NVP-Co10 and NVP-Co15 cathodes, which display capacity retention values of 78\%, 56\%, and 45\%, respectively. In Fig.~\ref{RATE}(c) we also present the long cyclic data measured on a different cell of NVP-Co5 electrode material, which shows an excellent recantation of 79.5\% after 1500 cycles. This demonstrates the excellent electrochemical reversibility of NVP-Co5 cathode, which reveals its potential to meet practical requirements for the promising fast-charging battery applications. The specific capacity showed a significant decline at 10 C rate for the NVP-Co10 and NVP-Co15 samples, indicating the sluggish kinetics for the higher concentrations of Co$^{3+}$, thereby lowering electrochemical performance. Therefore, the $x=$ 0.05 is considered to be the optimum sample for the overall charge transport of the electrode with lower polarization and ensuring fast reaction kinetics. Arumuguam {\it et al.} also reported excellent chemical, structural and dissolution stability of the Co-containing electrode as compared to nickel and manganese counterparts, owing to the large octahedral site as well as the stabilization energy \cite{Manthiram_Nature_Com_2020}. As a consequence, it is important to note here that the superior electrochemical performance of NVP-Co5 cathodes can be inferred as a cumulative effect of: (i) the Co$^{3+}$ ion forms CoO$_6$ octahedra by partially replacing VO$^{6}$ octahedra, thereby improving the structural stability of the NASICON framework; (ii) the doping creates high degree of local defects which enhances the intrinsic electronic conductivity from the viewpoint of defect chemistry and semiconductor theory \cite{Lv_CNM_2019, Li_AEM_2020, Xiao_SMALL_2023}; and (iii) the carbon coating on the particles also aids in improving the extrinsic electron conduction by providing active surface area and prevents direct contact of electrode particles with the electrolyte, thereby facilitating the swift ion mobility. Hence, an optimum concentration of dopants substantially improves electrode reaction kinetics with great structural stability and cycling performance. The obtained results are quite impressive in comparison to the values/performance reported in literature, as listed in Table.~\ref{Table-Comparison}.

\begin{table*}
     \caption{Comparison of the electrochemical performances (current rate, discharge capacity, number of cycles, capacity retention) of the NVP electrode modified with the element substitutions.}
\vskip 0.1 cm
    \centering
    \begin{tabular}{p{4.8cm} p{3.0cm} p{5.7cm} p{2cm}}
    \hline
    Materials&Synthesis Route&Electrochemical Performance&Reference\\
    \hline
Na$_{3.03}$V$_{1.97}$Ni$_{0.03}$(PO$_{4}$)$_{3}$/C&Sol-gel&1 C, 107.1 mAh g$^{-1}$, 100, 95.5\%&\cite{Li_AMI_2016}\\ 
Na$_{2.9}$V$_{1.98}$Mo$_{0.02}$(PO$_{4}$)$_{3}$/C&Solid-state&10 C,  90 mAh g$^{-1}$, 500, 83.5\%&\cite{Li_JMCA_2018}\\
Na$_{3+y}$V$_{2-y}$Mn$_{y}$(PO$_{4}$)$_{3}$/C&Sol-gel&5 C, 100 mAh g$^{-1}$, 199, 90\%&\cite{Ghosh_AES_2022}\\
Na$_{3}$Al$_{1/3}$V$_{5/3}$(PO$_{4}$)$_{3}$/C&Sol-gel&1 C, 100 mAh g$^{-1}$, 200, 95\%&\cite{Wu_AEM_2021}\\
Na$_{3}$V$_{1.9}$Cr$_{0.1}$(PO$_{4}$)$_{3}$&Sol-gel&2 C, 105 mAh g$^{-1}$, 40, 99\%&\cite{Aragon_CEC_2015}\\
  Na$_{3}$V$_{1.96}$Ce$_{0.04}$(PO$_{4}$)$_{3}$/C&Sol-gel&10 C, 90 mAh g$^{-1}$, 100, 99\%&\cite{Zheng_ECA_2017}\\
  Na$_{3.05}$V$_{1.95}$Mg$_{0.05}$(PO$_{4}$)$_{3}$/C&Sol-gel&10 C, 96.7 mAh g$^{-1}$, 180, 88.9\%&\cite{Li_CoM_2018}\\
  Na$_{2.9}$V$_{1.9}$Ti$_{0.1}$(PO$_{4}$)$_{3}$/C&Sol-gel&10 C, 106.3 mAh g$^{-1}$, 200, 95.8\%&\cite{Gu_CI_2019}\\
  Na$_{3}$V$_{1.7}$Mn$_{0.3}$(PO$_{4}$)$_{3}$/C&Sol-gel&2 C, 76 mAh g$^{-1}$, 30, 75\%&\cite{Klee_JPS_2016}\\	
Na$_{3}$V$_{1.95}$Co$_{0.05}$(PO$_{4}$)$_{3}$/C&Solid-state&0.1 C, 105 mAh g$^{-1}$ &This work\\
Na$_{3}$V$_{1.95}$Co$_{0.05}$(PO$_{4}$)$_{3}$/C&Solid-state&10 C, 80 mAh g$^{-1}$, 500, 92\%&This work\\
Na$_{3}$V$_{1.95}$Co$_{0.05}$(PO$_{4}$)$_{3}$/C&Solid-state&10 C, 80 mAh g$^{-1}$, 1500, 79.5\%&This work\\
\hline
 \hline
 \end{tabular}
 \label{Table-Comparison}
\end{table*}

In order to understand the sodium charge storage kinetics, we further perform cyclic voltammetry measurements at different scan rates of 0.1--1.0 mV/s in a potential window of 2.0--4.3 V. The cyclic voltammogram for all the samples show sharp redox peaks, as depicted in Figs.~S3(a--d), indicating reversible electrochemical reaction. The oxidation current starting at 3.47 V corresponds to the extraction of two Na ions from the Na2 site and the reduction current starting at 3.28 V reflects the insertion of two Na ions at the Na2 site \cite{Jian_AM_2017,Zheng_JEC_2018, Saravanan_AEM_2012}. In the NVP-Co5 sample, the cleavage of the reduction peaks represents the transfer of sodium ions from Na1 to Na2 site due to the local environment arrangement \cite{Jiang_JMCA_2016}. An additional peak corresponding to the activation of V$^{4+}$/V$^{5+}$ redox couple also appears at 3.95/3.89 V for all the samples, indicating the extraction of additional sodium at higher voltage, as seen clearly in the inset of Fig.~S3(a-d), which is in agreement with the charge-discharge profiles \cite{Wang_AMI_2020}. We observe the shift in oxidation peaks towards the higher potential and reduction peaks towards the lower potential with the increasing scan rate, as shown in Figs.~\ref{CV_CONTRIBUTION}(a, e) for the NVP and NVP-Co5 samples, respectively, which are associated with an increase in polarization at higher current density. Concurrently, the magnitudes of oxidation and reduction peaks also amplify as the scan rates rise, arising from the existence of electric double-layer capacitors and the capacitive current is in linear relationship with the scan rate \cite{Liu_Dalton_2022}, explained in detail in the below section. The peak currents at all the scan rates are calculated for the NVP and NVP-Co5 samples and their linear relationship with the (scan rate)$^{0.5}$, as shown in Figs.~\ref{CV_CONTRIBUTION}(b, f), respectively. The  overall charge storage reaction during Na insertion/extraction can be expressed as the following equations:
\begin{multline}
Na_{3}V_{2-x}^{3+}Co_{x}^{3+}(PO_{4})_{3} = Na_{1+x}V_{2-2x}^{4+}Co_{x}^{3+}(PO_{4})_{3} + \\(2-x)Na^{+} + (2-x)e^{-}
\end{multline}
\begin{multline}
 Na_{1+x}V_{2-2x}^{4+}Co_{x}^{3+}(PO_{4})_{3} =  Na_{1}V_{2-2x}^{4+}V_{x}^{5+}Co_{x}^{3+}(PO_{4})_{3} \\+ Na^{+} + xe^{-}
\end{multline}
 Considering that the current obeys power law relation, $i=a\nu$$^{b}$; where $a$ and $b$ are constants, $i$ and $\nu$ are the peak current and scan rate, respectively. The constant, $b$ is an indicator of the type of Na (de)intercalation process involved in the charge storage mechanism and the value of $b$ is calculated from the slope of the log(i$_{p}$) and log($\nu$) plot, as shown in Figs.~\ref{CV_CONTRIBUTION}(c, g) for the NVP and NVP-Co5 samples, respectively. As reported in the literature, if the value of $b=$ 0.5, the charge storage mechanism is termed as diffusion-controlled (faradaic) and the value of $b=$ 1.0 defines the charge storage mechanism as the surface-controlled (capacitive). However, if the value of $b$ lies between 0.5 and 1, charge storage mechanism is considered as a combination of both the above mentioned contributions and termed as pseudo-capacitive \cite{Mathis_AEM_2019}. Herein, the value of $b=$ 0.45 and  0.41 is calculated for the anodic peaks for both the NVP and NVP-Co5 samples; however, for the cathodic peak, the value of $b=$ 0.48 for the NVP, which increases to $b=$ 0.58 for the NVP-Co5 sample, signifying the involvement of both the bulk (phase change) and surface (adsoprtion on surface) intercalation of Na ions \cite{Babu_AEM_2020}. The contributions are quantified by the sum of capacitive (k$_{1}$$\nu$) and diffusion-controlled (k$_{2}$$\nu$$^{1/2}$) through the equation $i$(V) = k$_{1}$$\nu$ + k$_{2}$$\nu$$^{1/2}$ and values are depicted in Figs.~\ref{CV_CONTRIBUTION}(d, h). The pseudo-capacitive contribution rates of NVP-Co5 are higher than the NVP sample, which aids in the fast sodium storage and hence, improved capacity, especially at higher current rates \cite{Pu_AC_2021}. The apparent diffusion coefficient of sodium is determined from the Randles-Sevcik equation \cite{Li_CoM_2018}:
	\begin{equation}
		i_{p} = 2.69\times 10^{5}n^{3/2}ACD^{1/2}\nu^{1/2}
	\end{equation}
where, i$_{p}$ is the peak current (mA), $n$ is the number of electrons transferred in the redox reaction ($n=$ 2 in the present case), $A$ is the area of the electrode (1.33 cm$^{2}$), $C$ is the bulk concentration of the electrode (mol cm$^{-3}$), $D$ is the diffusion coefficient of the Na ions (cm$^{2}$ s$^{-1}$) and $\nu$ is the scan rate (mV s$^{-1}$). The calculated diffusion coefficient values, D$_{Na^{+}}$ are 6.4 $\times$ 10$^{-10}$ cm$^{2}$ s$^{-1}$ and 5.1 $\times$ 10$^{-10}$ cm$^{2}$ s$^{-1}$ for the anodic and cathodic peaks, respectively, for the NVP sample. For the NVP-Co5 cathode, we find the values of D$_{Na^{+}}$=3.5 $\times$ 10$^{-9}$ cm$^{2}$ s$^{-1}$  and 3.2 $\times$ 10$^{-9}$ cm$^{2}$ s$^{-1}$ for the anodic and cathodic peaks, respectively, describing the boost in the sodium ion migration dynamics with the incorporation of optimum amount of Co-substitution in the NVP sample. 
	
\begin{figure}
\includegraphics[width=\linewidth]{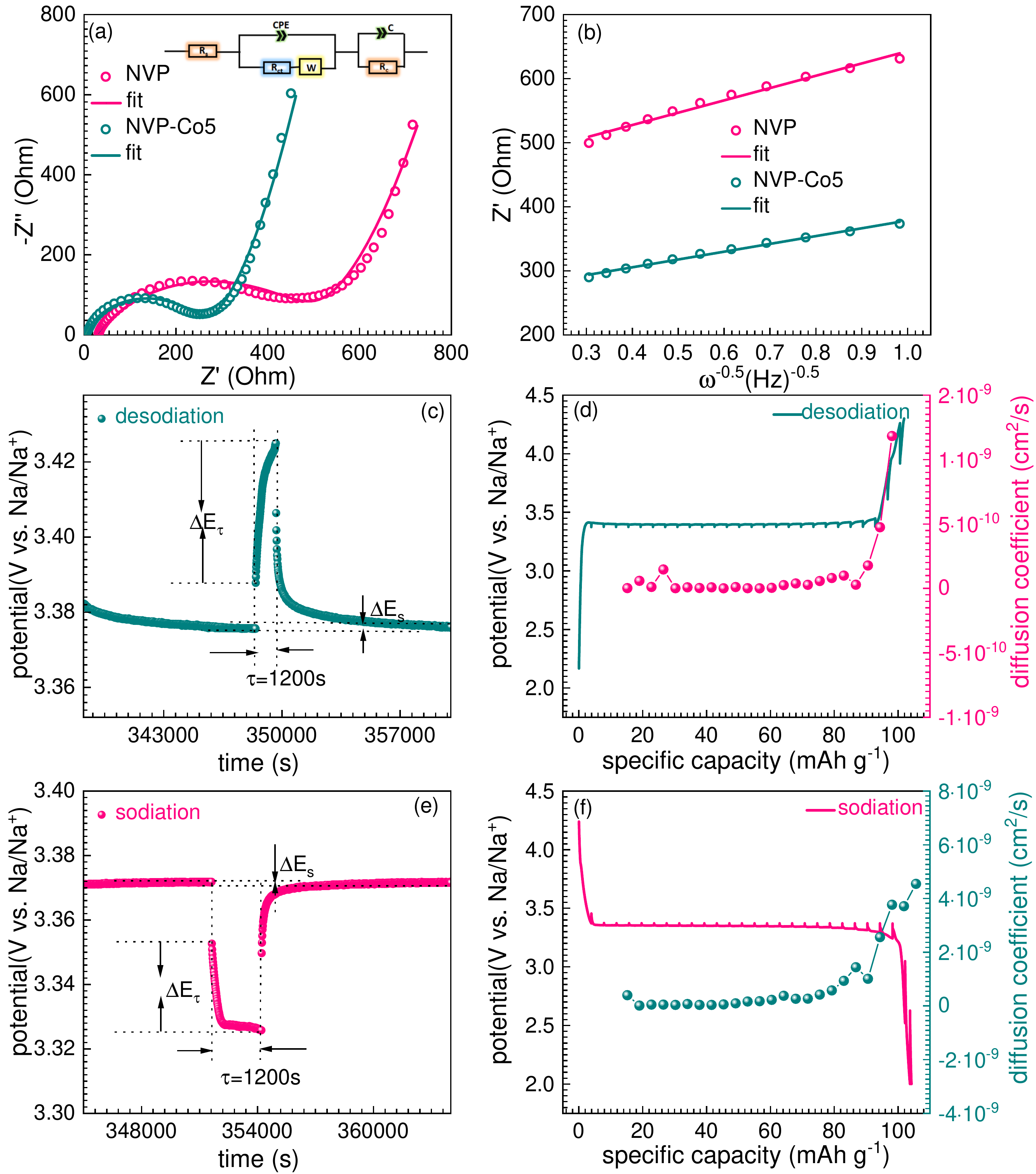}
\caption{The Nyquist Plot (Z' vs. - Z'') (a) and the Z' versus frequency plot (b) for the NVP and NVP-Co5 electrodes. The single titration curve for (c) charging and (e) discharging and the GITT profiles for the charge (c) and discharge (d) curves, along with their corresponding diffusion coefficients for the NVP-Co5 electrode.}
.\label{EIS-GITT}
\end{figure}
	
Furthermore, we use electrochemical impedance spectroscopy (EIS) technique to investigate the diffusion kinetics and resistance behavior at different interfaces. The cells were first cycled at 0.1~C rate for three cycles to stabilize the interface before recording the EIS spectrum. The Nyquist plot is presented in Fig.~\ref{EIS-GITT}(a), where the inset shows the analogous Randles's circuit model, which unveils the resistances involved in the migration of the sodium ions through the electrode-electrolyte interfaces. The intercept of the semicircle demonstrates the solution and contact resistance (R$_{s}$), and the diameter of the semicircle describes the charge transfer resistance (R$_{ct}$) across the electrode-electrolyte interface in the high-frequency region. The slope of the curve in the low-frequency region provides information about the diffusion of Na ions across the bulk of the electrode, represented by the Warburg diffusion component. The constant phase element (CPE) is incorporated for the polycrystalline and rough electrodes, corresponding to the double-layer capacitance at the solid electrolyte interphase, justifying that diffusion process is controlled by the interfacial and bulk charge transfer. Notably, the NVP-Co5 electrode exhibits the lower R$_{ct}$ value of 258 $\Omega$ as compared to  R$_{ct}=$ 424 $\Omega$ for the NVP electrode, which confirms that moderate doping of Co$^{3+}$ ions facilitates the easier charge migration and transport of electrons with the reduced interfacial resistance, which accounts for the superior rate capability and cycle life of the NVP-Co5 electrode. The apparent diffusion coefficient of the sodium ions across the electrode can be obtained from the Warburg impedances in the low-frequency region of the EIS spectra using the below equations 4 and 5 \cite{Rui_ECA_2010}:
\begin{equation}
D = \frac{R^{2}T^{2}}{2A^{2}\sigma^{2}n^{4}F^{4}C^{2}}
\end{equation}
\begin{equation}
Z_{re} = R_{s} + R_{ct} + \sigma\omega^{-0.5}
\end{equation}
where R is the gas constant, T is the temperature, A is the active surface area of the electrode (1.33 cm$^{2}$), F is the Faraday constant, $n$ is the number of electrons participating in the redox reaction (here 2 for two Na$^{+}$), C is the concentration of the sodium ions in the electrode (mol cm$^{-3}$) and $\sigma$ is the Warburg coefficient in the low-frequency region, related to the Z$_{re}$, in the above equation. The values of $\sigma$ are calculated from the linear fitting of the plot of Z$_{re}$ and reciprocal square root of frequency ($\omega^{-0.5}$), as depicted in Fig.~\ref{EIS-GITT}(b). The diffusion coefficient for the NVP cathode is calculated as 3.4 $\mathrm{\times}$ 10$^{-14}$ cm$^{2\ }$s${}^{-1}$ and the value of 8.5 $\mathrm{\times}$ 10$^{-14}$ cm$^{2}$ s$^{-1}$ is obtained for the NVP-Co5 cathode, indicating the enhanced electronic conduction and diffusion of Na ions across the electrode/electrolyte interface, which coincides well with the superior cycle and rate performance of the NVP-Co5 cathode. 
	
\begin{figure}
\includegraphics[width=\linewidth]{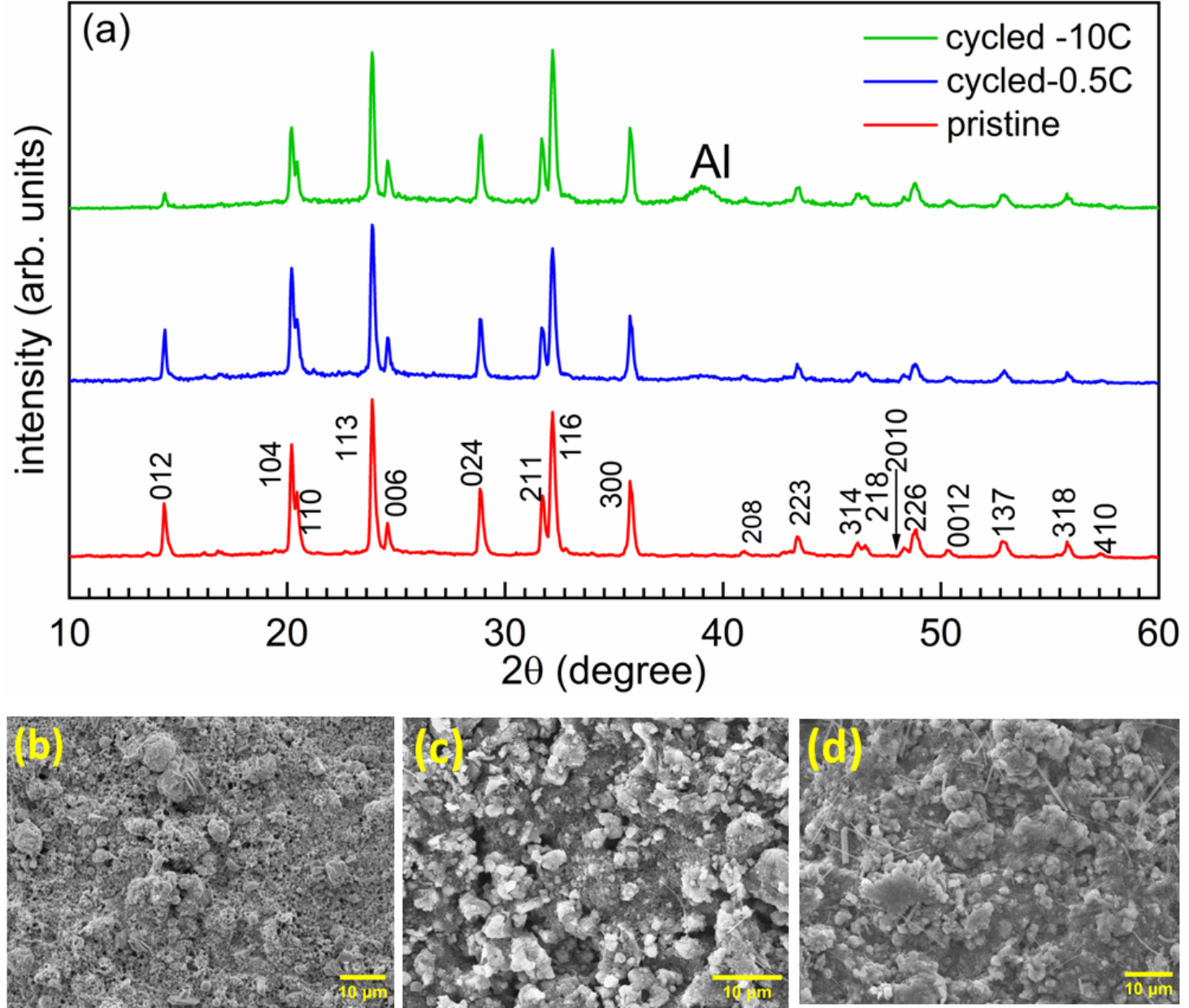}
\caption{The ex-situ XRD patterns in (a) and the SEM micrographs for the pristine electrode (b), the cycled electrode at 0.5 C rate for 200 cycles (c), and 10 C rate for 500 cycles (d) for the NVP-Co5 sample.}
\label{postmortem}
\end{figure}
	
To deeply explore the effect of substitution on the diffusion kinetics with the changes in the state-of-charge (SOC), the GITT experiments were conducted in the potential window of 2.0--4.3 V at a rate of 0.1 C. Initially, the half cell is activated by two charge-discharge cycles to ensure the stability of the electrode. The cell was first charged to the maximum cut-off voltage (4.3 V) with a current pulse ($\tau$) of 20 minutes, followed by a relaxation period of 180 minutes to ensure the homogeneous distribution of sodium ions and allow the system to reach electrochemical quasi-equilibrium OCV (dE/dt $\rightarrow$0 V/s). This whole experiment is repeated till the fully discharged voltage (2.0 V) is reached. The galvanostatic titration charge-discharge curves for the NVP-Co5 sample are depicted in Figs.~\ref{EIS-GITT}(c, e) where the voltage plateau appears at 3.4 V from the biphasic reaction. The sodium diffusion coefficient (D$_{Na^{+}}$) can be calculated using the Fick's second law \cite{Deiss_ECA_2005, Nickol_JES_2020, Zhu_JPCC_2010} as below:
\begin{equation}
D_{Na^{+}} =\frac{4}{\pi \tau }(\frac{m_{B}V_{M}}{M_{B}A})^2(\frac{{\Delta E}_S}{\tau \left(\frac{dE_{\tau }}{d\sqrt{\tau }}\right)})^2\ \ \ \ \ \ \ \tau \ll L^2/D_{{Na}^+}  
\end{equation}
The above equation is simplified into the following, considering the linear variation of the steady state voltage with the $\tau^{1/2}$ \cite{Deiss_ECA_2005, Nickol_JES_2020, Zhu_JPCC_2010}:
\begin{equation}
D_{Na^{+}} =\frac{4}{\pi \tau }(\frac{m_{B}V_M}{M_{B\ }A})^2(\frac{{\Delta E}_S}{\Delta E_{\tau }})^2 
\end{equation} 
where, the m$_{B}$ and M$_{B}$ are the active mass (mg) and molecular weight of the active material of the cathode (g/mol), V$_{M}$ is the molar volume (cm$^{3}$ mol$^{-1}$), L is the thickness ($\mu$m) and A is the active surface area of the electrode (cm$^{2}$), $\tau$ is the pulse time in which constant current is applied (s), $\Delta $E$_{S}$ represents the potential difference before and after the current pulse and $\Delta $E$_{\mathrm{\tau}}$ is the potential difference between the equilibrium potential and the potential maximum at the end of current pulse. During the two-phase reaction at the voltage platform (3.4 V), the diffusion is maintained in the range of 10$^{-11}$ cm$^{2}$/s while charging and discharging. However, in the solid solution region, at the start and the end of the charge-discharge curves, the diffusion coefficient varies with voltage in the range of 10$^{-9}$ cm$^{2}$ s$^{-1}$. The lower diffusion coefficient in the phase change of V$^{3+}$/V$^{4+}$ of the NVP-Co5 cathode arises from the resistance between the crystal framework and Na$^{+}$ ion, which restricts the movement of Na ions. This is found to be consistent with the voltage plateaus and redox peaks in the galvanostatic charge discharge and CV profiles, respectively \cite{Cheng_JPS_2020}.

Finally, we perform the postmortem analysis using the XRD and SEM to investigate the structural evolution in the NVP-Co5 electrode before and after cycling. For this purpose, the coin cells were dismantled, and the cycled electrode material was washed with dimethyl carbonate liquid electrolyte to remove the reactive species and then the electrodes were dried for 12 hrs in vacuum oven inside an Ar filled glovebox. The {\it ex-situ} XRD patterns of the electrode were carried out after testing at 0.5 C rate for 200 cycles and at 10 C rate for 500 cycles, as displayed in Fig.~\ref{postmortem}(a). The pristine diffraction peaks of the NVP-Co5 electrode match very well with the active material and the diffraction peak at 38$^{o}$ in the 10 C cycled electrode is related to the aluminium current collector. We find the stable structure ensuring excellent performance of the NVP-Co5 cathode. The morphology of the pristine and cycled NVP-Co5 electrodes is shown in Figs.~\ref{postmortem}(b--d). Also, there is no significant particle aggregation or cracking in the electrodes after long cycling apart from glass fiber separator seen on the surface, which are present due to the disassembling procedures. These results further support the consideration that the presence of cobalt in NVP electrode maintains structural coherence during the cycling process of sodium-ion batteries \cite{Wang_AMI_2020, Lim_JMCA_2014}.  \\
	
\section{\noindent ~Conclusions}

In summary, we successfully prepared a series of Na$_{3}$V$_{2-x}$Co$_{x}$(PO$_{4}$)$_{3}$/C ($x=$ 0, 0.05, 0.1, 0.15) cathode materials by a solid-state route. The Rietveld refinement of XRD patterns convey that Co doping does not affect the crystallographic structure, thereby, maintaining the three dimensional framework for the smooth migration of Na ions. Furthermore, the SEM and TEM measurements confirm the uniform carbon coating on the active material surface, which further contributed towards enhancing the extrinsic electrical conductivity. The valence state of V is confirmed to be 3+ from the XPS study.  Intriguingly, we find the best performance of $x=$ 0.05 sample, which shows the specific capacity of around 105 mAh g$^{-1}$ at 0.1 C rate. More importantly, we achieved the specific capacity of 80 mAh g$^{-1}$ at high current rate of 10 C after cycling for 500 cycles with about 92\% retention and nearly 100\% Coulombic efficiency. Through detailed analysis of galvanostatic intermittent titration and cyclic voltammetry data, the extracted diffusion coefficient of $x=$ 0.05 cathode is in the range of 10$^{-9}$--10$^{-11}$ cm$^{2}$ s$^{-1}$. The {\it ex-situ} XRD and SEM studies indicate the excellent structural and morphological stability after testing the material for 500 cycles at 10 C current rate. We demonstrate the excellent capacity retention (79.5\%) of optimally doped NVP-Co5 cathode upto 1500 cycles. 
	
\section{\noindent ~Conflict of interest }
The authors have no conflict of interest for this work. 
	
\section{\noindent ~Acknowledgements}
The financial support provided for this work by the National Science and Technology Council (NSTC) of Taiwan is gratefully appreciated. Simranjot K. Sapra would like to thank IIT Delhi and National Yang-Ming Chiao Tung University, Taiwan for the fellowship. We thank IIT Delhi for providing research facilities like  Raman, FTIR at the Central Research Facility and the Physics department for the XRD facility. We thank NSRRC Taiwan for providing beamtime for synchrotron XRD measurements. RSD acknowledges the Department of Science and Technology, Government of India for financial support through ``DST--IIT Delhi Energy Storage Platform on Batteries" (project no. DST/TMD/MECSP/2K17/07) and SERB--DST through a core research grant (file no.: CRG/2020/003436). R. S. Dhaka and J. -K. Chang thank NYCU and IIT Delhi for support through the MFIRP project (MI02683G).

\end{document}